\documentclass[journal]{IEEEtran}%导入的模板

\IEEEoverridecommandlockouts
\usepackage[utf8]{inputenc}
\usepackage{subfigure}

\usepackage{amsmath,amssymb,amsfonts}

\usepackage{mathtools}
\usepackage{amsmath} %\usepackage[cmex10]{amsmath}
\usepackage{amssymb}
\usepackage{amsfonts}
\usepackage{verbatim} %comment paragraph
\usepackage{bm,color,soul}
\usepackage{algorithm}
\usepackage{algorithmic} % algorithm
\usepackage{cite}
\usepackage{caption}
\usepackage{stfloats} % 
\usepackage[colorlinks, linkcolor=black, anchorcolor=black, citecolor=black]{hyperref} % superlink set
\usepackage{makecell} % Xhline Xcline
\newtheorem{thm}{Theorem}%类似于定义proposition
\newtheorem{prop}{Proposition}

\usepackage{array}
\usepackage{url}

\usepackage{balance}
\usepackage{flushend}

\makeatletter
\renewcommand{\maketag@@@}[1]{\hbox{\m@th\normalsize\normalfont#1}}%
\makeatother

\IEEEoverridecommandlockouts

\begin{document}
	%Integrated Sensing and Communications
\title{Joint Angle and Delay Cram\'{e}r-Rao Bound Optimization for ISAC}

\author{Chao Hu, Yuan Fang,~\IEEEmembership{Member,~IEEE} and Ling Qiu,~\IEEEmembership{Member,~IEEE}\vspace{-0.33in}

%\author{
%%	\IEEEauthorblockN{Chao Hu$^{1}$, Yuan Fang$^{2}$,  Ling Qiu$^{1*}$}
%	\IEEEauthorblockN{Chao Hu, Yuan Fang, and Ling Qiu, \textit{Member, IEEE}}
%	\IEEEauthorblockA{$^1$Key Laboratory of Wireless-Optical Communications, Chinese Academy of Sciences,  \\
%		School of Information Science and Technology,
%		University of Science and Technology of China,  Hefei, P. R. China. }
%	\IEEEauthorblockA{$^2$ Future Network of the Intelligence Institute (FNii) and School of Science and Engineering (SSE),\\
%		The Chinese University of Hong Kong (Shenzhen), Shenzhen, P. R. China.}
	% \IEEEauthorblockA{\{zhangsan\}@XXX.com, \{lisi, wangwu\}@XXX.edu.cn, {g.li}@XXX.com}
%	\IEEEauthorblockA
%	{$^1$\href{mailto:@XXX.edu.cn}
%		{sa21006191}@mail.ustc.edu.cn,
%		$^2$\href{mailto:g.li@XXX.com}{fangyuan}@cuhk.edu.cn,
%		$^1$\href{mailto:lisi@XXX.edu.cn}{lqiu}@ustc.edu.cn}  
%	{$^1${sa21006191}@mail.ustc.edu.cn,
%	$^2${fangyuan}@cuhk.edu.cn,
%	$^1${lqiu}@ustc.edu.cn} 
%\author{Chao Hu,}
%	{Yuan Fang,}
%	and {Ling Qiu,~\IEEEmembership{Member,~IEEE} 

%	\thanks{This work is supported by National Key R$\&$D Program of China (2022YFB2902302).
%	}
	\thanks{This work is supported by the National Key Research and Development Program of	China under Grant 2022YFB2902302.
    }
    \thanks{C. Hu and L. Qiu are with the Key Laboratory of Wireless Optical Communications, Chinese Academy of Sciences, School of Information Science and Technology, University of Science and Technology of China, Hefei 230052, China (e-mail: sa21006191@mail.ustc.edu.cn;
    	lqiu@ustc.edu.cn).}
    \thanks{Y. Fang is with the School of Science and Engineering (SSE) and the Future Network of Intelligence Institute (FNii), The Chinese University of Hong Kong, Shenzhen, Shenzhen 518172, China (e-mail:fangyuan@cuhk.edu.cn).}
	\thanks{Y. Fang and L. Qiu are corresponding authors.}
	%	\thanks{Chang Lu and Ling Qiu are with University of Science and Technology of China.}
	%\thanks{Yuan Fang is  with  The Chinese University of Hong Kong, Shenzhen.}
}		

	\maketitle \thispagestyle{empty}
	 \vspace{-0.1in}

\begin{abstract}
In this paper, we study a multi-input multi-output (MIMO) beamforming design in an integrated sensing and communication (ISAC) system, in which an ISAC base station (BS) is used to communicate with multiple downlink users and simultaneously the communication signals are reused for sensing multiple targets. Our interested sensing parameters are the angle and delay information of the targets, which can be used to locate these targets. Under this consideration, we first derive the Cram\'{e}r-Rao bound (CRB) for joint angle and delay estimation. Then, we optimize the transmit beamforming at the BS to minimize the CRB, subject to the communication rate requirement and the maximum transmit power constraint. In particular, we obtain the closed-form optimal solution in the case of single-target and single-user, and in the case of multi-target and multi-user scenario, the sparsity of the optimal solution is proven, leading to a reduction in computational complexity during optimization. The numerical results demonstrate that the optimized beamforming yields excellent positioning performance and effectively reduces the requirement for a large number of antennas at the BS.
\end{abstract}

% no keywords
%given by the dual-function radar-communication (DFRC) techniques
\vspace{-0.1in}
\section{Introduction}
Integrated Sensing and Communication (ISAC) technology is considered one of the promising key technologies for 6G, garnering significant attention from both academia and industry \cite{1}. By integrating sensing capabilities into the conventional cellular network, ISAC opens up numerous application possibilities, such as smart factories, drone monitoring, and intelligent transportation systems. These functionalities demand the system to possess the capability of actively locating targets, a task typically achieved using radar in the past.

Hence, it becomes imperative to investigate the positioning performance of ISAC. Compared with sensing-centric waveform design, communication signal-based waveform design can usually better meet the requirements of current high-speed communication, but it needs to maintain robust sensing performance \cite{2}, \cite{3}. To precisely and appropriately define the purpose of sensing, \cite{4}, \cite{5} introduced the Cramer-Rao bound (CRB) as a metric for sensing performance evaluation. The CRB enables the analysis of estimation performance for the required sensing parameters. However, their estimation approach often focuses on a single angle parameter or response matrix parameter, which fails to capture the intricate relationship between sensing and communication. Notably, a single parameter may inadequately represent the multifaceted impact of complex functions, such as positioning.
%Hence, it becomes imperative to investigate the positioning performance of ISAC. Previous studies have explored two waveform design schemes: sensing-centric and communication-centric \cite{2}. However, the former often falls short in providing the high communication rate necessary for current high data rate transmissions. On the other hand, the communication signal-based waveform design scheme \cite{3} seeks to achieve a high communication rate while maintaining robust sensing performance.
%
%To precisely and appropriately define the purpose of sensing, [4] and [5] introduced the Cram\'{e}r-Rao bound (CRB) as a metric for sensing performance evaluation. The CRB enables the analysis of estimation performance for the required sensing parameters. However, their estimation approach often focuses on a single angle parameter or response matrix parameter, which fails to capture the intricate relationship between sensing and communication. Notably, a single parameter may inadequately represent the multifaceted impact of complex functions, such as positioning.

Recently, \cite{6} investigated a multi-base station cooperative positioning ISAC system that locates target through angle and delay estimation due to their excellent positioning performance. However, the sensing metric based on these estimations is still rarely considered in ISAC systems. Although a few articles like \cite{7}, \cite{8} explored location-centered sensing metrics, they focused on radar-communication coexistence systems, relying on existing radar systems, besides, the former merely considered the estimation of angle for simplifying formula which is not enough to achieve location, the latter only analysed the performance of bound, but did not fully research the resource allocation. It is worth noting that while \cite{6} first adopts the joint CRB of angle and delay in ISAC and accurately describes the positioning performance, it only considers a single antenna transmitter and studies energy allocation. In reality, current base stations (BSs) typically employ multiple antennas, and distributed structures impose high synchronization requirements on the system. Therefore, this paper addresses these challenges by minimizing the CRB for estimated angle and delay under rate and energy constraints, based on a multi-antenna monostatic BS system.

Specifically, we first derive the CRB for joint angle and delay estimation, and then formulate an beamforming design problem to minimize the CRB performance, while ensuring the communication rate requirement and the maximum power constraint. In the single-target and single-user scenario, we obtain a closed-form optimal solution based on orthogonal projection. In the case of multi-target and multi-user, we prove the sparsity of the solution, which extremely simplifys computational complexity. Simulations are conducted to demonstrate the accuracy and effectiveness of our analysis and optimization, and it reveals insightful results that is deserved to be further exploited.

%{\it Notations:} 
%Vectors and matrices are denoted by bold lower and upper-case letters, respectively. $\boldsymbol{I}$ represents an identity matrix with appropriate dimensions. For a square matrix $\boldsymbol{A}$, $\operatorname{Tr}(\boldsymbol{A})$ denotes
%its trace, and $\boldsymbol{A} \succeq 0$ means that A is positive semidefinite.
%For an arbitrary-size matrix $\boldsymbol{B}$, $\operatorname{vec}\left(\boldsymbol{B}\right)$, $\operatorname{rank}\left(\boldsymbol{B}\right)$, $\boldsymbol{B}^{T}$, and $\boldsymbol{B}^{H}$ denote
%its vectorization, rank, transpose, and conjugate transpose, respectively. The Hadamard (element-wise) and Kronecker products are denoted by $\odot$ and $\otimes$. $\operatorname{Re}(\cdot)$ denotes the real  part of the argument, $\mathbb{E}(\cdot)$ represents the expectation of random variables, $\|\cdot\|$ denotes the
%Euclidean norm of a vector, while $\operatorname{diag}(\cdot)$ outputs a vector containing the diagonal entries of the input matrix. 
\vspace{-0.1cm}
	
	\section{System Model}%系统模型***************************************

%\begin{figure}[!h]
%	\centering
%	\includegraphics[width=0.35\textwidth]{fig0.png}
%	\caption{The ISAC model.}\label{figmode0}
%\end{figure}

\begin{figure}[t]
	%\begin{figure}[!h]
	\vspace{-0.0cm}
	\centering
%	    \vspace{-0.4cm}	
	%    \vspace{-0.4cm}	
	\includegraphics[width=0.4\textwidth]{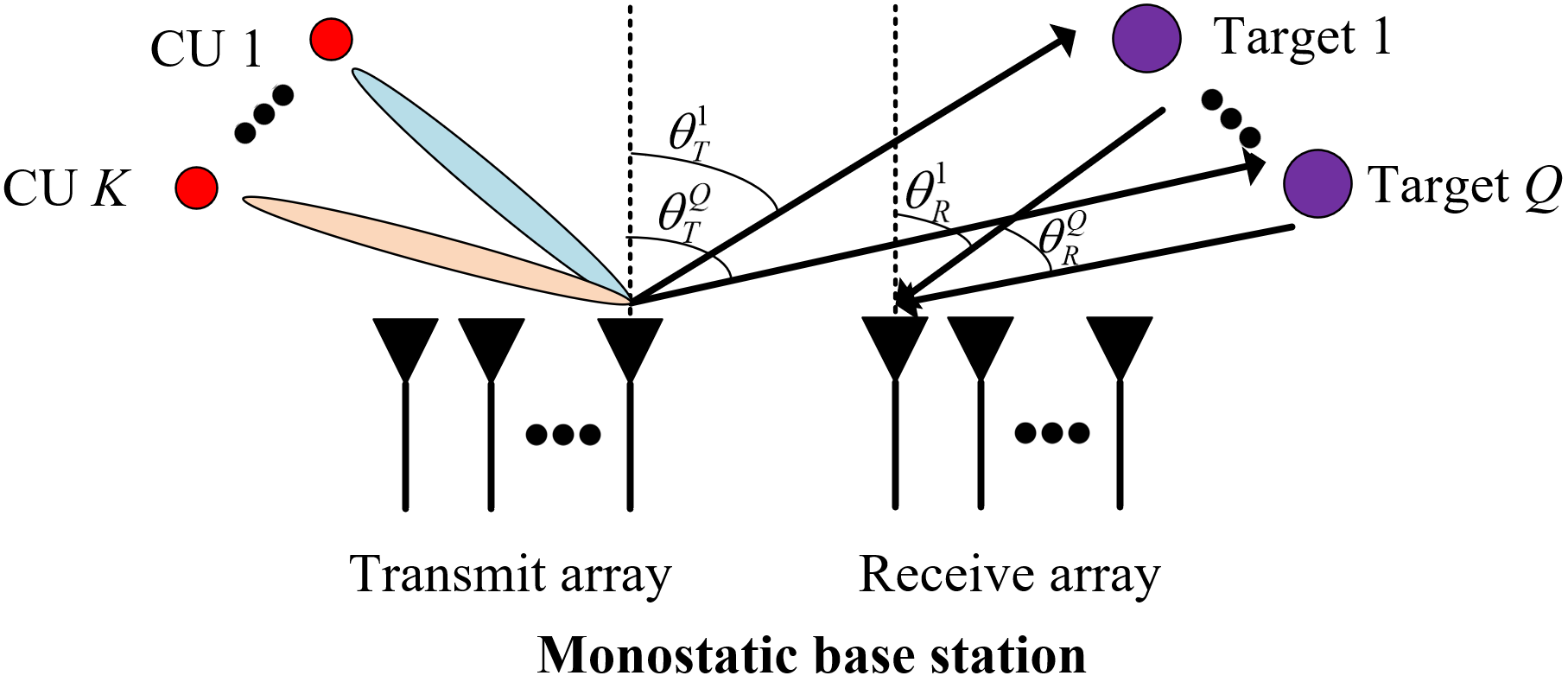}
	\captionsetup{font={small}}
	\caption{An illustration of the system.}\label{figmode0}
	\vspace{-0.8cm}	
\end{figure}

In this paper, we consider an ISAC system in which a BS is equipped with uniform linear array (ULA), containing $N_t$ transmit antennas and $N_r$ receive antennas ($N_t \leq N_r$). The whole downlink communication block is leveraged for performing the dual tasks of multiple radar targets localization and communication data transmission, as depicted in Fig. \ref{figmode0} where the angle of departure is assumed being equal to the angle of arrival \cite{4}. Let $\mathcal{K} = \left\{1, \cdots, K\right\}$ denote the set of communication users (CUs), and  $\mathcal{Q} = \left\{1, \cdots, Q\right\}$ denote that of sensing targets. The coordinates of the BS and the $q$-th target are denoted as $({x}_{0}, {y}_{0})$ and $({x}_q, {y}_q)$, respectively, $q \in \mathcal{Q}$. 

For communication link, let $s_k(t)$   and $\boldsymbol{h}_{k} \in \mathbb{C}^{{{N_t}} \times 1}$ denote the transmit signal and the channel vector from the BS to CU $k\in \mathcal K$, respectively, the former is a random variable with zero mean and unit variance, and the latter is assumed to be quasi-static and estimated perfectly.  Then the received signal at CU $k$ is expressed as
\begin{align} \label{ReceSig}
	y_k(t) = \boldsymbol{h}_{k}^H \boldsymbol{w}_k s_k(t) + \!\sum_{l \neq k, l \in \mathcal K}\! \boldsymbol{h}_{k}^H \boldsymbol{w}_l s_l(t) + n_k(t),
\end{align}
where $\boldsymbol{w}_{k} \in \mathbb{C}^{{{N_t}} \times 1}$ is the beamforming vector of the $k$-th CU, $n_k(t) \sim \mathcal{CN}(0,\sigma_k^2)$ denotes the noise at the receiver of CU $k$, which means that $n_k(t)$ is a circularly symmetric complex Gaussian (CSCG) random variable with zero mean and variance $\sigma_k^2$. The corresponding achievable rate of $k$-th CU is given by $R = \log_2(1+\gamma_k)$
where $	{\rm \gamma}_{k}= \frac{\left| \boldsymbol{h}_{k}^H \boldsymbol{w}_k\right|^2 }{ \sum\limits_{l \neq k, l \in \mathcal K} \lvert \boldsymbol{h}_{k}^H \boldsymbol{w}_l \rvert^2   + \sigma_k^2},  \forall  k \in  \mathcal{K}$. 
%\begin{align}  \label{SINR}
%	{\rm \gamma}_{k}= \frac{\left| \boldsymbol{h}_{k}^H \boldsymbol{w}_k\right|^2 }{ \sum\limits_{l \neq k, l \in \mathcal K} \lvert \boldsymbol{h}_{k}^H \boldsymbol{w}_l \rvert^2   + \sigma_k^2},  \forall  k \in  \mathcal{K}.
%\end{align}

Next, we consider the sensing link, in which  the communication signals $s_{k}(t)$'s are reused for  sensing. Suppose that the radar processing is implemented over an interval $\mathcal{T}$ with duration $T$, which is sufficiently long so that $\int_{\mathcal{T}}\lvert s_{k}(t)\rvert^2dt = T \cdot \mathbb{E}\left[\lvert s_{k}(t)\rvert^2\right] = T$ and $ \int_{\mathcal{T}} s_{k}(t)s^*_{l}(t)dt=T \cdot  \mathbb{E}\left[s_{k}(t)s^*_{l}(t)\right] = 0$, $\forall k \neq l$ \cite{6}. Then the reflected echoes received at the BS can be given in the form
\begin{align} \label{ReflectedRadar}
	\boldsymbol{r}(t) = \sum_{q=1}^Q \sum_{k=1}^K \alpha_{q} \boldsymbol{a}(\theta_q) \boldsymbol{b}(\theta_q)^T \boldsymbol{w}_k s_k(t-\tau_{q}) + \boldsymbol{n}_{r}(t),
\end{align}
where $\boldsymbol{n}_{r}(t) \sim \mathcal{C N}(\boldsymbol{0}, \sigma^{2}\boldsymbol{I}_{N_r})$ denotes the complex additive white Gaussian noise, $\alpha_{q}$ is a complex amplitude capturing the effects of the target radar cross section (RCS) and the path loss for the radar propagation path between the BS and target $q$. $\theta_q$ is the azimuth angle of the $q$-th target relative to the BS, $\boldsymbol{b}(\theta_q) \in \mathbb{C}^{{{N_t}} \times 1}$ and $\boldsymbol{a}(\theta_q) \in \mathbb{C}^{{{N_r}} \times 1}$ are transmit and receive steering vectors of the antenna array of the BS. Here, we choose the center of the ULA antennas as the reference point, which means the transmit steering vector and its derivative are expressed as ${\mathbf{b}}\left( \theta  \right) = {\left[ {{e^{ - j\frac{{{N_t-1}}}{2}\pi \sin \theta }},{e^{ - j\frac{{{N_t} - 3}}{2}\pi \sin \theta }}, \ldots ,{e^{j\frac{{{N_t-1}}}{2}\pi \sin \theta }}} \right]^T}$, ${\mathbf{\dot b}}\left( \theta  \right) = {\left[ { - j{b_1}\frac{{{N_t-1}}}{2}\pi \cos \theta , \ldots ,j{b_{{N_t}}}\frac{{{N_t-1}}}{2}\pi \cos \theta } \right]^T}$,
%\begin{equation}\label{eq20}
%	{\mathbf{b}}\left( \theta  \right) = {\left[ {{e^{ - j\frac{{{N_t-1}}}{2}\pi \sin \theta }},{e^{ - j\frac{{{N_t} - 3}}{2}\pi \sin \theta }}, \ldots ,{e^{j\frac{{{N_t-1}}}{2}\pi \sin \theta }}} \right]^T},
%\end{equation}
%\begin{equation}\label{eq21}
%	{\mathbf{\dot b}}\left( \theta  \right) = {\left[ { - j{a_1}\frac{{{N_t-1}}}{2}\pi \cos \theta , \ldots ,j{a_{{N_t}}}\frac{{{N_t-1}}}{2}\pi \cos \theta } \right]^T},
%\end{equation}		
where $b_i$ represents the $i$-th entry of ${\mathbf{b}}\left( \theta  \right)$. The receive steering vector and its derivative have similar form. According to the symmetry, it can be easily verified that $	{{\mathbf{b}}^H}{\mathbf{\dot b}} = 0,{{\mathbf{a}}^H}{\mathbf{\dot a}} = 0, \forall \theta$,
%\begin{equation}\label{eq22}
%	{{\mathbf{b}}^H}{\mathbf{\dot b}} = 0,{{\mathbf{a}}^H}{\mathbf{\dot a}} = 0, \forall \theta,
%\end{equation}
where ${{\mathbf{b}}}$, ${{\mathbf{a}}}$, ${{\mathbf{\dot b}}}$, and ${{\mathbf{\dot a}}}$ denote ${{\mathbf{b}}} \left(\theta\right)$, ${{\mathbf{a}}}\left(\theta\right)$, ${{\mathbf{\dot b}}}\left(\theta\right)$, and ${{\mathbf{\dot a}}}\left(\theta\right)$, respectively. $\tau_{q}$ denotes the propagation delay associated with $k$-th signal and $q$-th target, and can be expressed as $\tau_{q} = \frac{\sqrt{({x}_q-{x}_0)^2\!+({y}_q-{y}_0)^2}}{2c}   \triangleq \frac{R_{q}}{2c}$,
%\begin{align} \label{Delay}
%	\tau_{q} = \frac{\sqrt{({x}_q-{x}_0)^2\!+({y}_q-{y}_0)^2}}{2c}   \triangleq \frac{R_{q}}{2c},
%\end{align}
where $c$ is the speed of electromagnetic wave.
%	\vspace{-0.3cm}
%\section{Estimation CRB Derivation}

Based on the received echo signal in (\ref {ReflectedRadar}), we then derive the joint CRB of angle and delay in the general case. Denote the $ 2Q \times 1 $ vector of unknown target parameters as $\boldsymbol{\varsigma} \triangleq\left[\boldsymbol{\theta}^{T},\boldsymbol{\tau}^{T}\right]^{T} 
$, with $\boldsymbol{\theta} \triangleq\left[\theta_{1}, \ldots, \theta_{Q}\right]^{T}$ and $\boldsymbol{\tau} \triangleq\left[\tau_{1}, \ldots, \tau_{Q}\right]^{T} $. For an unbiased estimator $\hat{\boldsymbol{\varsigma}}$, the covariance matrix is lower bounded as 	$\boldsymbol{C}_{\hat{\boldsymbol{\varsigma}}}=\mathbb{E}\left\{(\hat{\boldsymbol{\varsigma}}-\boldsymbol{\varsigma})(\hat{\boldsymbol{\varsigma}}-\boldsymbol{\varsigma})^{H}\right\} \geq \boldsymbol{C}_{\mathrm{CRB}}$. 
%With $\boldsymbol{n_r} \sim \mathcal{C N}(\boldsymbol{0}, \sigma^{2}\boldsymbol{I}_{N_r})$ in (\ref{ReflectedRadar}), we can express the $ 2Q \times 2Q $ Fisher information matrix (FIM) in terms of the parameter submatrices 
%\begin{equation}
%\begin{aligned} 
%	\boldsymbol{F}=\left[\begin{array}{ll}
%	\boldsymbol{F}_{\boldsymbol{\theta}, \boldsymbol{\theta}} & \boldsymbol{F}_{\boldsymbol{\theta}, \boldsymbol{\tau}} \\
%	\boldsymbol{F}_{\boldsymbol{\tau}, \boldsymbol{\theta}} & \boldsymbol{F}_{\boldsymbol{\tau},\boldsymbol{\tau}}.
%\end{array}\right] 
%\end{aligned}
%\end{equation}
%The following Theorem 1 derives each submatrix term.
%Based on the definition of FIM in (8) and the received echo signal at the BS, we can obtain the FIM for the parameters estimation as shown in the following theorem. 
Based on the received echo signal at the BS, the Fisher information matrix (FIM) $\boldsymbol{F}$ is given in Theorem \ref {CRLB}, and the estimation CRB is calculated by $\boldsymbol{C}_{\mathrm{CRB}} = \boldsymbol{F}^{-1}$\cite{10}.
\begin{thm}\label{CRLB}
%For the signal model in (\ref{ReflectedRadar}), the submatrices of FIM in (\ref{CRLB_All}) are
The FIM $\boldsymbol{F}$ for estimating $\boldsymbol{\varsigma}$ is given by
\begin{small} 
\begin{equation}\label{CRLB_All}
	\boldsymbol{F}=\left[\begin{array}{ll}
		\boldsymbol{F}_{\boldsymbol{\theta}, \boldsymbol{\theta}} & \boldsymbol{F}_{\boldsymbol{\theta}, \boldsymbol{\tau}} \\
		\boldsymbol{F}_{\boldsymbol{\tau}, \boldsymbol{\theta}} & \boldsymbol{F}_{\boldsymbol{\tau}, \boldsymbol{\tau}} 
	\end{array}\right]=\frac{2}{\sigma^{2}} \operatorname{Re}\left[\begin{array}{ll}
		\boldsymbol{F}_{\boldsymbol{\theta}, \boldsymbol{\theta}}^{\prime} & \boldsymbol{F}_{\boldsymbol{\theta}, \tau}^{\prime} \\
		\boldsymbol{F}_{\boldsymbol{\tau}, \boldsymbol{\theta}}^{\prime} & \boldsymbol{F}_{\boldsymbol{\tau}, \boldsymbol{\tau}}^{\prime}
	\end{array}\right],
\end{equation}
\end{small} 
where
%\begin{equation}\label{CRLB_All_2}
%	\begin{aligned} 
	%		\boldsymbol{F}=\frac{2}{\sigma^{2}} \operatorname{Re}\left[\begin{array}{ll}
		%			\boldsymbol{F}^{'}_{\boldsymbol{\theta}, \boldsymbol{\theta}} & \boldsymbol{F}^{'}_{\boldsymbol{\theta}, \boldsymbol{\tau}} \\
		%			\boldsymbol{F}^{'}_{\boldsymbol{\tau}, \boldsymbol{\theta}} & \boldsymbol{F}^{'}_{\boldsymbol{\tau}, \boldsymbol{\tau}}
		%		\end{array}\right] ,
	%	\end{aligned}
%\end{equation}
%where	
%\begin{equation}
%	\begin{aligned} 
	\begin{small} 
		\begin{align} \label{FIM_theta_prime} 
		\boldsymbol{F}^{\prime}_{\boldsymbol{\theta}, \boldsymbol{\theta}}&=\left(\dot{\boldsymbol{A}}^{H} \dot{\boldsymbol{A}}\right) \odot\left(\boldsymbol{C}^{*} \boldsymbol{B}^{H}\left(\boldsymbol{W} \boldsymbol{s} \boldsymbol{s}^{H} \boldsymbol{W}^{H}\right)^{*} \boldsymbol{B} \boldsymbol{C}\right)\nonumber\\ 
		&+\left(\dot{\boldsymbol{A}}^{H} \boldsymbol{A}\right) \odot\left(\boldsymbol{C}^{*} \boldsymbol{B}^{H}\left(\boldsymbol{W} \boldsymbol{s s}^{H} \boldsymbol{W}^{H}\right)^{*} \dot{\boldsymbol{B}} \boldsymbol{C}\right) \nonumber\\ 
		&+\left(\boldsymbol{A}^{H} \dot{\boldsymbol{A}}\right) \odot\left(\boldsymbol{C}^{*} \dot{\boldsymbol{B}}^{H}\left(\boldsymbol{W} \boldsymbol{s} \boldsymbol{s}^{H} \boldsymbol{W}^{H}\right)^{*} \boldsymbol{B C}\right)\nonumber\\ 
		&+\left(\boldsymbol{A}^{H} \boldsymbol{A}\right) \odot\left(\boldsymbol{C}^{*} \dot{\boldsymbol{B}}^{H}\left(\boldsymbol{W} \boldsymbol{s} \boldsymbol{s}^{H} \boldsymbol{W}^{H}\right)^{*} \dot{\boldsymbol{B}} \boldsymbol{C}\right), 
		\end{align}
	\end{small} 
\vspace{-0.1cm}
\begin{small} 
	\begin{align} \label{FIM_thetatau_prime}
		\boldsymbol{F}^{\prime}_{\boldsymbol{\theta}, \boldsymbol{\tau}}
		&=\!\boldsymbol{F}^{\prime T}_{\boldsymbol{\tau}, \boldsymbol{\theta}}\nonumber\\
		&=\!\left(\!\dot{\boldsymbol{A}}^{H} \boldsymbol{A}\!\right) \odot\left(\!\boldsymbol{C}^{\boldsymbol{*}} \boldsymbol{B}^{H}\left(\boldsymbol{W s s}^{H} \boldsymbol{W}^{H}\right)^{\boldsymbol{*}} \boldsymbol{B C}(-j \omega)\!\right)\nonumber \\
		&+\!\left(\!\boldsymbol{A}^{H} \boldsymbol{A}\!\right) \odot\left(\!\boldsymbol{C}^{*} \boldsymbol{B}^{H}\left(\boldsymbol{W} \boldsymbol{s} \boldsymbol{s}^{H} \boldsymbol{W}^{H}\right)^{*} \boldsymbol{B C}(-j \omega)\!\right), 		
	\end{align}
\end{small} 
%\end{equation}
\begin{small} 
\begin{equation}
	\begin{aligned} \label{FIM_tautau_prime}
		\boldsymbol{F}^{\prime}_{\boldsymbol{\tau}, \boldsymbol{\tau}}
		=\left(\boldsymbol{A}^{H} \boldsymbol{A}\right) \odot\left(\boldsymbol{C}^{\boldsymbol{*}} \boldsymbol{B}^{H}\left(\boldsymbol{W} \boldsymbol{s} \boldsymbol{s}^{H} \boldsymbol{W}^{H}\right)^{*} \boldsymbol{B} \boldsymbol{C}\right).
	\end{aligned}
\end{equation}
\end{small} 
\end{thm}
\begin{IEEEproof}
See Appendix \ref{CRLB_Derivation}.
\end{IEEEproof}
\vspace{-0.1cm}
\section{Transmit Beamforming Design}
%Because we are estimating a vector of parameters, which results in CRLB being a matrix. In such a situation, to obtain a scalar objective function, \cite{11} introduced several CRLB-based scalar metrics for optimization. In this paper, we choose the objective functionas $\operatorname{Tr}(\boldsymbol{C}_{\rm{CRLB}})$ based on an average sense or the A-optimality criterion. The A-optimality criterion has a low computational complexity for optimization \cite{11} and directly targets the minimization of the parameters of interest. Hence, we consider the following CRLB approximation-based optimization problem
In this section, we form the optimization problem\footnote{What we considered is one time slot among the entile process with the previous information of angles known, so we employ the angle information as well as the delay in last slot to complete the optimization, which is based on a assumption that the target is not moving with a superhigh speed. For the initial value of parameters, it is necessary to complete the estimation of parameters which can refer to \cite{henk}.} that takes  CRB and communication rate as the performance matrics.  Because we are estimating a vector of parameters, the CRB is a matrix. To obtain a scalar objective function, \cite{11} introduced several CRB-based scalar metrics for optimization. In this paper, we choose the objective function $\operatorname{Tr}(\boldsymbol{C}_{\rm{CRB}})$ where $\operatorname{Tr}\left(\cdot\right)$ denote the calculation for trace, aimed to balance the units used for different target parameters via an average sense, which yieds the following optimization problem
\begin{subequations} \label {s_0}
\begin{align} 
	\min_{\left\{\boldsymbol{w}_k\right\}_{k=1}^{K}} \quad &\operatorname{Tr}(\boldsymbol{C}_{\rm{CRB}}) \\
\text {s.t.}\quad &\log _{2}\left(1+\gamma_{k}\right) \geq \Gamma_{k}, \forall  k \in  \mathcal{K} \text {, } \\
&\sum\nolimits_{k=1}^K \|\boldsymbol{w}_k\|^{2} \leq P_{T} \text {,} 
\end{align}
\end{subequations}
where $\Gamma_{k}$ denotes the rate threshold of $k$-th CU, $P_{T}$ is the transmit power budget, and $\|\cdot\|$ denotes the
Euclidean norm.
Next, we analyze the problem in two cases, and give some insights about the solution to the problem.
\vspace{-0.1in}
\subsection{Single-Target and Single-User Case}
% From [, Proposition1], $\operatorname{Tr}(\boldsymbol{F}^{-1}_{\boldsymbol{\theta}, \boldsymbol{\theta}}) + \operatorname{Tr}(\boldsymbol{F}^{-1}_{\boldsymbol{\tau},\boldsymbol{\tau}})$ is both a lower bound and an approximation of $\operatorname{Tr}(\boldsymbol{C}_{\rm{CRLB}})$.

%\frac{2|\alpha|^{2}}{\sigma^{2}}\left(\frac{1}{\|\dot{\boldsymbol{a}}\|^{2}\left|\boldsymbol{b}^{H} \boldsymbol{w}_1\right|^{2}\!+\!\|\boldsymbol{a}\|^{2}\left|\dot{\boldsymbol{b}}^{H} \boldsymbol{w}_1\right|^{2}}\!\right.\notag\\
%&\left.+\frac{1}{\omega^{2}\|\boldsymbol{a}\|^{2}\left|\boldsymbol{b}^{H} \boldsymbol{w}_1\right|^{2}}\right) \\  %, let $\sigma_{c}^2$, $\alpha$ be the communication noise power and complex amplitude
 In this subsection, we consider the case of one CU and one target. Taking advavtage of Theorem \ref{CRLB} in this scenario, the $\operatorname{Tr}(\boldsymbol{C}_{\rm{CRB}})$ is equal to $\frac{1}{F_{\theta,\theta}}+\frac{1}{F_{\tau,\tau}}$ on account of $F_{\tau,\theta}=F_{\theta,\tau}=0$ with the fact\footnote{These can be obtained through Slepian-Bangs formula or Theorem \ref {CRLB} with the properties ${{\mathbf{b}}^H}{\mathbf{\dot b}} = 0,{{\mathbf{a}}^H}{\mathbf{\dot a}} = 0, \forall \theta.$} that $F_{\theta,\theta}$ and $F_{\tau,\tau}$ are $	\frac{2|\alpha_1|^2}{\sigma^2}\big(\|\dot{\boldsymbol{a}}\|^2|\boldsymbol{b}^H\boldsymbol{w}_{1}|^2+\|\boldsymbol{a}\|^2|\dot{\boldsymbol{b}}^H\boldsymbol{w}_{1}|^2\big)$ and $	\frac{2\omega^2|\alpha_1|^2}{\sigma^2}(\|\boldsymbol a\|^2|\boldsymbol b^H\boldsymbol{w}_{1}|^2)$. Then, problem (\ref {s_0}) is transformed into  
\begin{subequations}\label{s_1}
\begin{align} 
	\min _{\boldsymbol{w}_1}\quad &f_0(\boldsymbol{w}_1)\\
\text {s.t.}\quad&\left|\boldsymbol{h}_{1}^{H} \boldsymbol{w}_1\right|^{2} \geq\left(2^{\Gamma_{1}}-1\right) \sigma_{1}^{2} \text {, } \label{sinr1}\\
&\|\boldsymbol{w}_1\|^{2} \leq P_{T}, 
\end{align}
\end{subequations}
where $f_0(\boldsymbol{w}_1)=\frac{1}{F_{\theta,\theta}}+\frac{1}{F_{\tau,\tau}}$. Although problem (\ref{s_1}) is non-convex and intractable due to the fractional objective function, we give a closed-form optimal solution to that in the following theorem via the orthogonal projection.
\begin{thm}\label{S_D}
Define $\boldsymbol{d}_{x}=\frac{\boldsymbol{b}}{\|\boldsymbol{b}\|}$, $\boldsymbol{d}_{y}=\frac{\dot{\boldsymbol{b}}}{\| \dot{\boldsymbol{b}}\|}$, and $\boldsymbol{d}_{z}=\frac{\boldsymbol{h}_{1}-\boldsymbol{d}_{x}^{H} \boldsymbol{h}_{1} \boldsymbol{d}_{x}-\boldsymbol{d}_{y}^{H} \boldsymbol{h}_{1} \boldsymbol{d}_{y}}{\left\|\boldsymbol{h}_{1}-\boldsymbol{d}_{x}^{H} \boldsymbol{h}_{1} \boldsymbol{d}_{x}-\boldsymbol{d}_{y}^{H} \boldsymbol{h}_{1} \boldsymbol{d}_{y}\right\|}$, the optimal solution $\boldsymbol{w}_{1}^{*}$ to problem (\ref{s_1}) is
\begin{align}
	 \boldsymbol{w}_{1}^{*}=x_{1}\boldsymbol{d}_{x}+x_{2}\boldsymbol{d}_{y}+x_{3}\boldsymbol{d}_{z},
\end{align}
where $x_{1}$, $x_{2}$, $x_{3}$ are given systematically as follows.
\begin{itemize}
	\item { Case 1: When $\|\dot{\boldsymbol{a}}\|^{2}\|\boldsymbol{b}\|^{2}-\|\boldsymbol{a}\|^{2}\|\dot{\boldsymbol{b}}\|^{2} \frac{\left|\boldsymbol{h}_{1}^{H} \boldsymbol{d}_{z}\right|^{2}-\left|\boldsymbol{h}_{1}^{H} \boldsymbol{d}_{x}\right|^{2}}{\left|\boldsymbol{h}_{1}^{H} \boldsymbol{d}_{z}\right|^{2}-\left|\boldsymbol{h}_{1}^{H} \boldsymbol{d}_{y}\right|^{2}} \geq 0$, we have $ x_{1} = \sqrt{\frac{\left|\boldsymbol{h}_{1}^{H} \boldsymbol{d}_{z}\right|^{2} P_{T}-\left(2^{\Gamma_{1}}-1\right) \sigma_{1}^{2}}{\left|\boldsymbol{h}_{1}^{H} \boldsymbol{d}_{z}\right|^{2}-\left|\boldsymbol{h}_{1}^{H} \boldsymbol{d}_{x}\right|^{2}}}$, $x_{2}=\sqrt{\frac{\left|\boldsymbol{h}_{1}^{H}\boldsymbol{d}_{z}\right|^{2}P_{T}-\left(2^{{\Gamma_{1}}}-1\right)\sigma_{1}^{2}}{\left|\boldsymbol{h}_{1}^{H}\boldsymbol{d}_{z}\right|^{2}-\left|\boldsymbol{h}_{1}^{H}\boldsymbol{d}_{y}\right|^{2}}-\frac{\left|\boldsymbol{h}_{1}^{H}\boldsymbol{d}_{z}\right|^{2}-\left|\boldsymbol{h}_{1}^{H}\boldsymbol{d}_{x}\right|^{2}}{\left|\boldsymbol{h}_{1}^{H}\boldsymbol{d}_{z}\right|^{2}-\left|\boldsymbol{h}_{1}^{H}\boldsymbol{d}_{y}\right|^{2}}\left|x_{1}\right|^{2}} $, and $x_{3} =\sqrt{ P_T-\left|x_{1}\right|^{2}-\left|x_{2}\right|^{2}}$.
	}
	\item { Case 2: When $\|\dot{\boldsymbol{a}}\|^{2}\|\boldsymbol{b}\|^{2}-\|\boldsymbol{a}\|^{2}\|\dot{\boldsymbol{b}}\|^{2} \frac{\left|\boldsymbol{h}_{1}^{H} \boldsymbol{d}_{z}\right|^{2}-\left|\boldsymbol{h}_{1}^{H} \boldsymbol{d}_{x}\right|^{2}}{\left|\boldsymbol{h}_{1}^{H} \boldsymbol{d}_{z}\right|^{2}-\left|\boldsymbol{h}_{1}^{H} \boldsymbol{d}_{y}\right|^{2}} < 0$ and $\omega^{2}\|\boldsymbol{a}\|^{2}\left\|\boldsymbol{b}\right\|^{2}\left|x_{1}\right|^{2}=\|\boldsymbol{a}\|^{2}\left\|\dot{\boldsymbol{b}}\right\|^{2}\frac{\left|\boldsymbol{h}_{1}^{H}\boldsymbol{d}_{z}\right|^{2}-\left|\boldsymbol{h}_{1}^{H}\boldsymbol{d}_{x}\right|^{2}}{\left|\boldsymbol{h}_{1}^{H}\boldsymbol{d}_{z}\right|^{2}-\left|\boldsymbol{h}_{1}^{H}\boldsymbol{d}_{y}\right|^{2}}-\|\dot{\boldsymbol{a}}\|^{2}\left\|\boldsymbol{b}\right\|^{2}$, we have 
		\begin{align}
			x_{1}=\left\{\begin{array}{l}{x_{1,\max},x_{1,0}> x_{1,\max}},\\ {x_{1,0},x_{1,0}\leq x_{1,\max},}\\ \end{array}\right. 
		\end{align}
		%			\begin{equation}\label{w1_close}
			%				\begin{aligned}
				%					\boldsymbol{w}_{1}=\left\{\begin{array}{l}{x_{1}\boldsymbol{d}_{x}+x_{3}\boldsymbol{d}_{z},\|\dot{\boldsymbol{a}}\|^{2}+\omega^{2}\|\boldsymbol{a}\|^{2}>\|\boldsymbol{a}\|},\\ {c_0\boldsymbol{d}_{x}+x_{2}\boldsymbol{d}_{y}+x_{3}\boldsymbol{d}_{z},\|\dot{\boldsymbol{a}}\|^{2}+\omega^{2}\|\boldsymbol{a}\|^{2}\leq\|\boldsymbol{a}\|,}\\ \end{array}\right. \\
				%				\end{aligned}
			%			\end{equation}	
		%			\begin{align}
			%				x_{1}=\left\{\begin{array}{l}{x_{1}\boldsymbol{d}_{x}+x_{3}\boldsymbol{d}_{z},x_{1,0} > x_{1,\max},\\ {c_0\boldsymbol{d}_{x}+x_{2}\boldsymbol{d}_{y}+x_{3}\boldsymbol{d}_{z},x_{1,0}\leq x_{1,\max},}\\ \end{array}, 
				%			\end{align}	
			where $ x_{1,\max} = \sqrt{\frac{\left|\boldsymbol{h}_{1}^{H} \boldsymbol{d}_{z}\right|^{2} P_{T}-\left(2^{\Gamma_{1}}-1\right) \sigma_{1}^{2}}{\left|\boldsymbol{h}_{1}^{H} \boldsymbol{d}_{z}\right|^{2}-\left|\boldsymbol{h}_{1}^{H} \boldsymbol{d}_{x}\right|^{2}}}$ and $x_{1,0} = \sqrt{ \frac{\beta_2\lambda_1}{2(\beta_2\lambda_2-\beta_1)}}$. $\beta_1,\beta_2,\beta_3,\lambda_1$ and $\lambda_2$ are defined in the proof process aimed to simplify expatiatory formulas. $x_{2}$ and $x_{3}$ can be acquired as case 1, and the same is true below.
		}
		\item { Case 3: When $\|\dot{\boldsymbol{a}}\|^{2}\|\boldsymbol{b}\|^{2}-\|\boldsymbol{a}\|^{2}\|\dot{\boldsymbol{b}}\|^{2} \frac{\left|\boldsymbol{h}_{1}^{H} \boldsymbol{d}_{z}\right|^{2}-\left|\boldsymbol{h}_{1}^{H} \boldsymbol{d}_{x}\right|^{2}}{\left|\boldsymbol{h}_{1}^{H} \boldsymbol{d}_{z}\right|^{2}-\left|\boldsymbol{h}_{1}^{H} \boldsymbol{d}_{y}\right|^{2}} < 0$ and $\omega^{2}\|\boldsymbol{a}\|^{2}\left\|\boldsymbol{b}\right\|^{2}\left|x_{1}\right|^{2}>\|\boldsymbol{a}\|^{2}\left\|\dot{\boldsymbol{b}}\right\|^{2}\frac{\left|\boldsymbol{h}_{1}^{H}\boldsymbol{d}_{z}\right|^{2}-\left|\boldsymbol{h}_{1}^{H}\boldsymbol{d}_{x}\right|^{2}}{\left|\boldsymbol{h}_{1}^{H}\boldsymbol{d}_{z}\right|^{2}-\left|\boldsymbol{h}_{1}^{H}\boldsymbol{d}_{y}\right|^{2}}-\|\dot{\boldsymbol{a}}\|^{2}\left\|\boldsymbol{b}\right\|^{2}$, we have
			\begin{align}
				x_{1}=\left\{\begin{array}{l}{x_{1,\max},x_{1,\mathrm{rig1}}> x_{1,\max}},\\ {x_{1,\mathrm{rig1}},x_{1,\mathrm{rig1}}\leq x_{1,\max},}\\ \end{array}\right. 
			\end{align}
			where $ x_{1,\mathrm{rig1}} = \sqrt{\frac{\beta_2\lambda_1\left(\sqrt{(\beta_2\lambda_2-\beta_1)\beta_3}+\beta_1-\beta_2\lambda_2\right)}{\left(\beta_2\lambda_2-\beta_1\right)\left(\beta_3+\beta_1-\beta_2\lambda_2\right)}}$.
		}
		\item { Case 4: When $\|\dot{\boldsymbol{a}}\|^{2}\|\boldsymbol{b}\|^{2}-\|\boldsymbol{a}\|^{2}\|\dot{\boldsymbol{b}}\|^{2} \frac{\left|\boldsymbol{h}_{1}^{H} \boldsymbol{d}_{z}\right|^{2}-\left|\boldsymbol{h}_{1}^{H} \boldsymbol{d}_{x}\right|^{2}}{\left|\boldsymbol{h}_{1}^{H} \boldsymbol{d}_{z}\right|^{2}-\left|\boldsymbol{h}_{1}^{H} \boldsymbol{d}_{y}\right|^{2}} < 0$ and $\omega^{2}\|\boldsymbol{a}\|^{2}\left\|\boldsymbol{b}\right\|^{2}\left|x_{1}\right|^{2}<\|\boldsymbol{a}\|^{2}\left\|\dot{\boldsymbol{b}}\right\|^{2}\frac{\left|\boldsymbol{h}_{1}^{H}\boldsymbol{d}_{z}\right|^{2}-\left|\boldsymbol{h}_{1}^{H}\boldsymbol{d}_{x}\right|^{2}}{\left|\boldsymbol{h}_{1}^{H}\boldsymbol{d}_{z}\right|^{2}-\left|\boldsymbol{h}_{1}^{H}\boldsymbol{d}_{y}\right|^{2}}-\|\dot{\boldsymbol{a}}\|^{2}\left\|\boldsymbol{b}\right\|^{2}$, we have
			\begin{align}
				x_{1}=\left\{\begin{array}{l}{x_{1,\max},x_{1,\mathrm{lef2}}\geq x_{1,\max}},\\ {x_{1,\mathrm{lef2}},x_{1,\mathrm{lef2}}\leq x_{1,\max}\leq x_{1,\mathrm{rig2}},}\\ \end{array}\right. 
			\end{align}	
			where $x_{1,\mathrm{lef2}} =\sqrt{\frac{\beta_2\lambda_1\left(\sqrt{(\beta_2\lambda_2-\beta_1)\beta_3}+\beta_1-\beta_2\lambda_2\right)}{\left(\beta_2\lambda_2-\beta_1\right)\left(\beta_3+\beta_1-\beta_2\lambda_2\right)}}$ and $x_{1,\mathrm{rig2}} = \sqrt{\frac{\beta_2\lambda_1\left(-\sqrt{(\beta_2\lambda_2-\beta_1)\beta_3}+\beta_1-\beta_2\lambda_2\right)}{\left(\beta_2\lambda_2-\beta_1\right)\left(\beta_3+\beta_1-\beta_2\lambda_2\right)}}$.
		}
	\end{itemize}
\end{thm}
\begin{IEEEproof}
	See Appendix \ref{S_Derivation}.
\end{IEEEproof}
\vspace{-0.1in}
\subsection{Multi-Target and Multi-User Case}
We first analyze the relationship between the elements of the FIM and $\boldsymbol{w}_k$. Based on $\boldsymbol{r}_{1}=\boldsymbol{A C B}{ }^{T} \boldsymbol{W} \boldsymbol{s}+\boldsymbol{n}_{1}$, which is defined in Appendix \ref{CRLB_Derivation}, we let $\boldsymbol{D}=\boldsymbol{A C B}{ }^{T}$, and assume that $\dot{\boldsymbol{D}}_m$ is the derivative with respect to $\boldsymbol{\varsigma}_{m}$, then
%\begin{equation}
\begin{align} \label{CRLB_mm}
[\boldsymbol{F}]_{m n}&=2 \operatorname{Re}\left\{\operatorname{Tr}\left(\frac{\partial \boldsymbol{\mu}^{H}}{\partial \boldsymbol{\varsigma}_{m}} \boldsymbol{R}^{-1} \frac{\partial \boldsymbol{\mu}}{\partial \boldsymbol{\varsigma}_{n}}\right)\right\} \notag\\
%&=\frac{2}{\sigma^{2}} \operatorname{Re}\operatorname{Tr}\left((\dot{\boldsymbol{D}}_m \boldsymbol{W} \boldsymbol{s})^{H}(\dot{\boldsymbol{D}}_n \boldsymbol{W} \boldsymbol{s})\right) \\
%&=\frac{2}{\sigma^{2}} \operatorname{Re}\left\{\operatorname{Tr}\left(\boldsymbol{s}^{H} \boldsymbol{W}^{H} \dot{\boldsymbol{D}}_m^{H} \dot{\boldsymbol{D}}_n \boldsymbol{W} \boldsymbol{s}\right)\right\} \\
%&=\frac{2}{\sigma^{2}} \operatorname{Re}\left\{\operatorname{Tr}\left(\boldsymbol{s s}^{H} \boldsymbol{W}^{H} \dot{\boldsymbol{D}}_m^{H} \dot{\boldsymbol{D}}_n \boldsymbol{W}\right)\right\} \\
&=\frac{2}{\sigma^{2}} \operatorname{Re}\left\{\boldsymbol{w}_{v}^{H}\left(\left(\boldsymbol{s}^{*} \boldsymbol{s}^{T}\right) \otimes \dot{\boldsymbol{D}}_m^{H} \dot{\boldsymbol{D}}_n\right) \boldsymbol{w}_{v}\right\}, 
\end{align}
%\end{equation}
where $\boldsymbol{w}_{v}=\operatorname{vec}(\boldsymbol{W})$, $\operatorname{vec}\left(\cdot\right)$ denote the vectorization and $\otimes$ represents the Kronecker product. Thus, the optimization problem is formulated as
\begin{subequations}\label{MM_OP_1}
\begin{align} 
	\min _{\boldsymbol{w}_{v}} \quad &\operatorname{Tr}(\boldsymbol{C}_{\rm{CRB}}) \\
\text {s.t.}\quad &\frac{\left|\boldsymbol{g}_{k}^{H} \boldsymbol{w}_{v}\right|^{2}}{\boldsymbol{w}_{v}^{H} \boldsymbol{L}_{k} \boldsymbol{w}_{v}+\sigma_{k}^{2}} \geq \left(2^{\Gamma_{k}}-1\right), \forall k \in \mathcal{K}, \label {ctb}\\
&\|\boldsymbol{w}_v\|^{2} \leq P_{T}, 
\end{align}
\end{subequations}
where $\boldsymbol{g}_{k}= 
\boldsymbol{e}_{k} \otimes \boldsymbol{h}_{k}$ and $\boldsymbol{L}_{k}=\left(\boldsymbol{1}-\boldsymbol{e}_{k}\right)\left(\boldsymbol{1}-\boldsymbol{e}_{k}\right)^{T} \otimes \boldsymbol{h}_{k} \boldsymbol{h}_{k}^{H}$ with $\boldsymbol{e}_{k}$ denoting the $k$-th column of the identity matrix in relevant dimension and $\boldsymbol{1}$ representing a vector in which all elements are one. Note that problem (\ref{MM_OP_1}) is non-convex due to the objective function and communication  rate constraints  in (\ref{ctb}) are non-convex. Based on SDR technique, by letting $\boldsymbol{W}_{v}=\boldsymbol{w}_{v} \boldsymbol{w}_{v}^{H}$ and dropping the rank-one constraint of $W_v$, problem (\ref{MM_OP_1}) is relaxed as
\begin{subequations}\label{MM_OP_SDP}
	\begin{align} 
	\min_{\boldsymbol{W}_{v},\left\{t_{q}\right\}_{q=1}^{2 Q}}  & \boldsymbol{1}^{T} \boldsymbol{t} \\
\text {s.t.}\quad & \left[\begin{array}{ll}
	\boldsymbol{F} & \boldsymbol{e}_{q} \\
	\boldsymbol{e}_{q}^T & t_{q}
\end{array}\right] \succeq 0, q = 1,\dots,2Q, \\
&\frac{\operatorname{Tr}\left(\boldsymbol{g}_{k} \boldsymbol{g}_{k}^{H} \boldsymbol{W}_{v}\right)}{\operatorname{Tr}\left(\boldsymbol{L}_{k} \boldsymbol{W}_{v}\right)+\sigma_{k}^{2}} \geq\left(2^{\Gamma_{k}}-1\right), \forall k \in \mathcal{K}, \\
&\operatorname{Tr}\left(\boldsymbol{W}_{v}\right) \leq P_{T}, 
	\end{align}
\end{subequations}
where $\left\{t_{q}\right\}_{q=1}^{2 Q}$ is a the introduced auxiliary variables. Note that $\boldsymbol{F}$ is a linear function of $\boldsymbol{W}_{v}$, which
implies that (\ref{MM_OP_SDP}) is a convex semidefinite program (SDP) and thus can be solved via standard convex optimization tools \cite{CVX}. Furthermore, we present the following proposition to demonstrate the structure of the optimal solution to problem (\ref{MM_OP_SDP}).
\begin{prop}\label{M_D} 
The optimal $\boldsymbol{W}^{*}_{v}$ obtained to (\ref{MM_OP_SDP}) is expressed as
\begin{align}\label{M_W}
	\boldsymbol{W}_{v}^{*}=\boldsymbol{U}_{T} \boldsymbol{\Lambda} \boldsymbol{U}_{T}^{H},
\end{align}	
where $\boldsymbol{U}_{T} \triangleq[\boldsymbol{B}, \dot{\boldsymbol{B}}]^{*}$, and $\boldsymbol{\Lambda} \in \mathbb{C}^{2 Q \times 2 Q} $ is a positive semidefinite matrix.
\end{prop}
\begin{IEEEproof}
	See Appendix \ref{M_Derivation}.
\end{IEEEproof}
Note that Proposition 1 gives a low computational complexity method to obtain the optimal solution to problem (\ref{MM_OP_SDP}). In particular, the optimization in problem (\ref{MM_OP_SDP}) can equivalently be performed over
$\boldsymbol{\Lambda}$ instead of $\boldsymbol{W}_{v}$ so that the optimization dimension reduces from $N_r$ to $2Q$ for usually $N_r \gg Q$. After obtaining the optimal solution to problem (\ref{MM_OP_SDP}), we can acquire the rank-one solution to problem (\ref{MM_OP_1}) through Gaussian randomization or eigenvalue decomposition \cite{luo}.
\vspace{-0.15in}
\section{Numerical Results}

%\begin{figure*}[htbp] %通栏
%	\begin{minipage}[t]{0.33\linewidth} %调节两个子图左右间距
%		\centering
%		\includegraphics[width=2.5in, height=1.9in]{7_25_CRB_VS_Pt_Theta_AM_MRT.eps} %调节单个子图大小
%		\captionsetup{font={small}}
%		\caption{The CRLB of $\theta$ versus $P_T$} %子图下标题
%		\label{fig2} %引用标签
%	\end{minipage}%
%	\begin{minipage}[t]{0.33\linewidth}
%		\centering
%		\includegraphics[width=2.5in, height=1.9in]{7_25_CRB_VS_Pt_Delay_AM_MRT.eps}
%		\captionsetup{font={small}}
%		\caption{The CRLB of $\tau$ versus $P_T$}
%		\label{fig3}
%	\end{minipage}%
%	\begin{minipage}[t]{0.33\linewidth}
%		\centering
%		\includegraphics[width=2.5in, height=1.9in]{7_25_CRB_vs_P_LA_15degree_MRT_u.eps}
%		\captionsetup{font={small}}
%		\caption{The CRLB of La versus $P_T$}
%		\label{fig4}
%	\end{minipage}
%\end{figure*}

In this section, the numerical results are provided to clarify the previous analysis and the validity of beamforming. We set the BS equipped with $N_t = 16$ and $N_r = 20$ antennas, serving $K = 3$  CUs and locating $Q = 3$ radar targets. The interval between adjacent antennas of the BS is half-wavelength. Carrier frequency is set as 6 GHz \cite{6}. The angles of the CUs and targets are set in the range $[-\frac{\pi}{2}, \frac{\pi}{2}]$. The overall power budget is $P_T = 20$ dBm and  the noise variances are set as $\sigma_k^2 = -90$ dBm, $\sigma^2 = -10$ dBm. For single-target and single-user case, the line of sight (LoS) model is considered for the channel from the BS to the CU with the path loss -70 dB. For multi-target and multi-user case, the communication channel is set as Rayleigh fading like \cite{4}. 

\begin{figure}[!h]
%\begin{figure}[b]
	\centering
    \vspace{-0.4cm}	
	\includegraphics[width=0.35\textwidth]{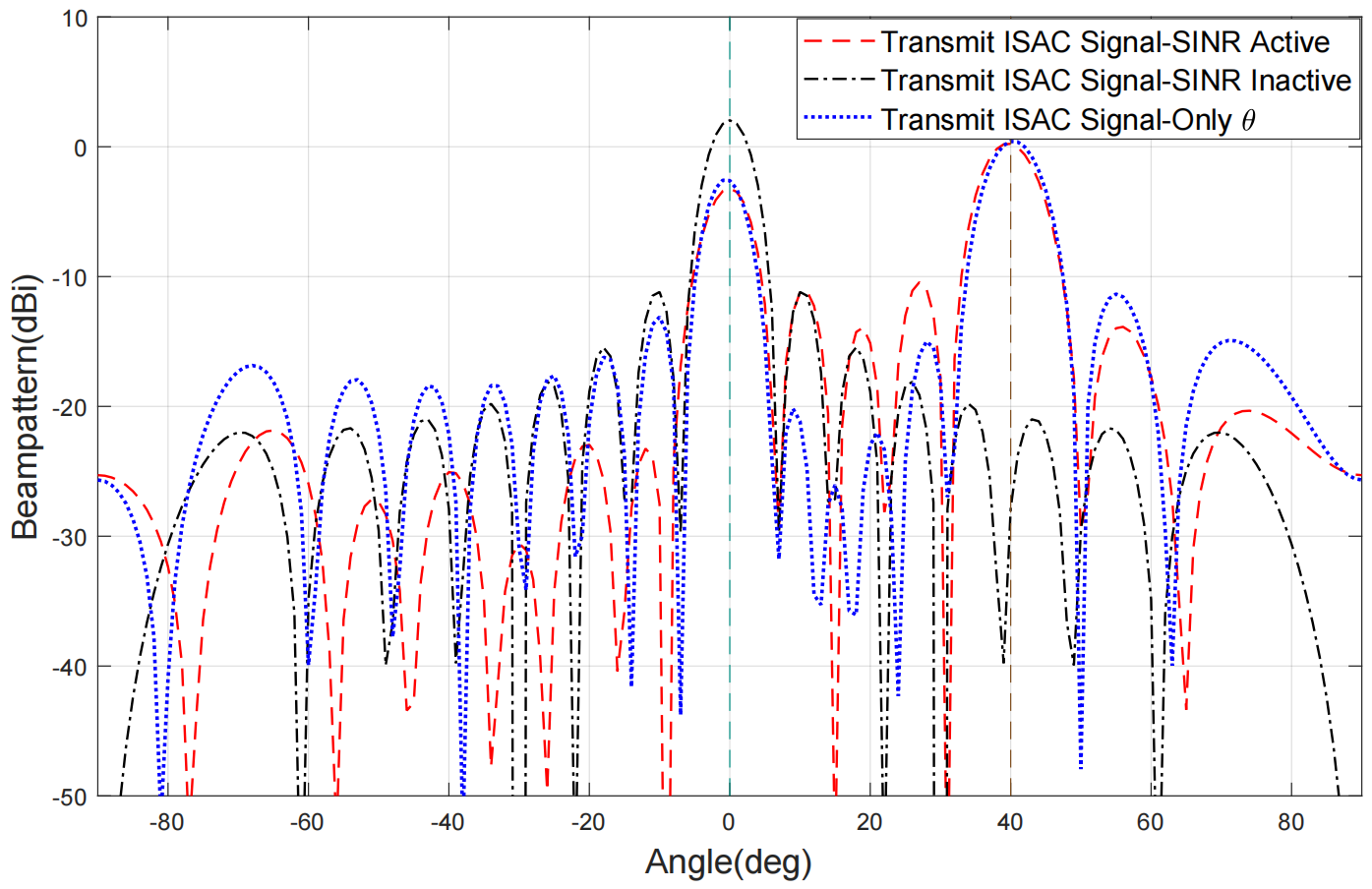}
	\captionsetup{font={footnotesize}}
	\caption{The transmit beampattern for single-target and single-user case, where the target and CU are located respectively at azimuth angles $0^{\circ}$ and $40^{\circ}$.}\label{figmodel}
	\vspace{-0.4cm}	
\end{figure}

Fig. \ref{figmodel} shows the resultant beampatterns where $\Gamma_{1} = 7.5 $ bps/Hz. When the communication rate constraint can not be satisfied which means (\ref{s_1}) is not feasible, the energy is no longer allocated to the CU, and the best sensing performance can be obtained, apparently it is not the optimal solution to problem (\ref{s_1}), but a forced scheme when the communication rate is not capable to be fulfilled. It is shown that not only does the optimal closed-form solution acquire a good sensing performance, it also guarantees the communication performance, even though its sensing performance is slightly lower than inactive's. Compared with the scheme that uses the term “only $\theta$” to represent the beamforming in \cite{4}, the optimal closed-form solution obtained through the method in this paper almost coincides with it in Fig. \ref{figmodel}, which indicates that the proposed method in this paper takes into account the angle estimation performance while ensuring the delay estimation performance.

\begin{figure*}[htbp]
	\vspace{-0.2cm}	
	\centering
	\captionsetup{font={footnotesize}}
	\subfigure[CRB of $\theta$ versus $P_T$.]{
		\begin{minipage}[t]{0.32\linewidth}\label{f1}
			\centering
			\includegraphics[width=2.5in]{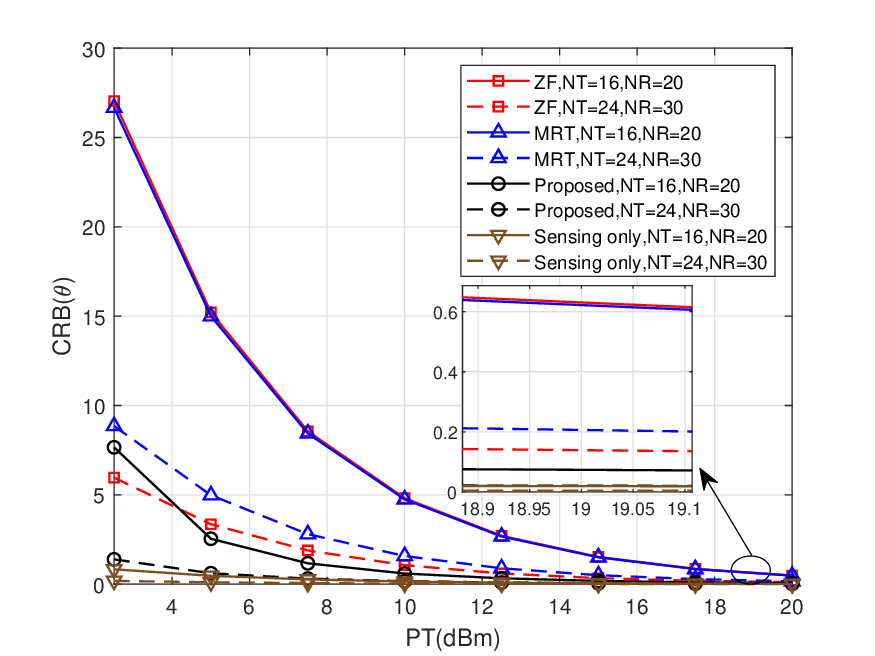}
		\end{minipage}
	}%
	\subfigure[CRB of $\tau$ versus $P_T$.]{
		\begin{minipage}[t]{0.31\linewidth}\label{f2}
			\centering
			\includegraphics[width=2.5in]{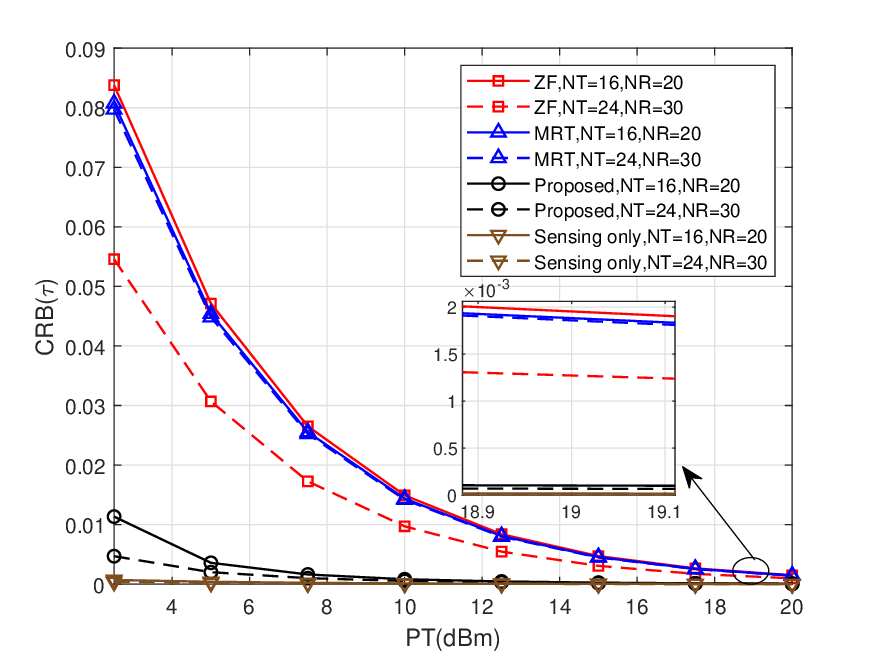}
		\end{minipage}
	}		
	\subfigure[CRB of positioning versus $P_T$.]{
		\begin{minipage}[t]{0.31\linewidth}\label{f3}
			\centering
			\includegraphics[width=2.5in]{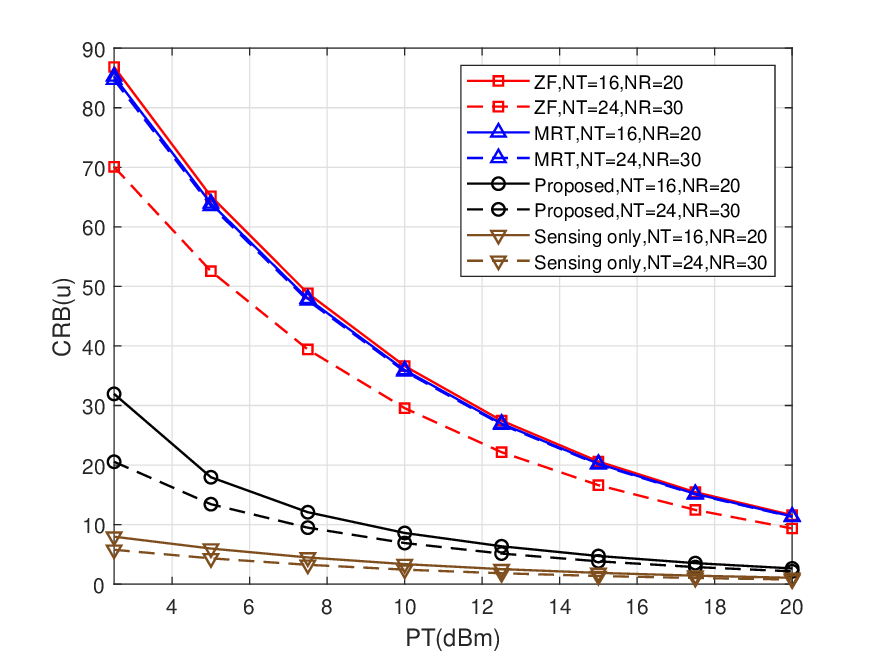}
		\end{minipage}
	}%
	\centering
	\caption{The estimation CRB versus $P_T$.}
	
	\label{f30}
	\vspace{-0.4cm}	
\end{figure*}
Fig. \ref{f30} show the relationship between CRB of the estimated parameters and total transmitted energy, where “CRB(u)” represents the lower bound on the estimate of the Cartesian coordinates of the targets, and its definition is as follows: $\rm{CRB}(u) = [\operatorname{Tr}(\widetilde{\boldsymbol{J}}^{-1})]^{1/2}$. $\widetilde{\rm\boldsymbol{J}} = (\frac{\partial\boldsymbol{\varsigma}}{\partial\boldsymbol{u}})\boldsymbol{F}(\frac{\partial\boldsymbol{\varsigma}}{\partial\boldsymbol{u}})^{T}$ is derived by the chain rule, and $\boldsymbol{u}$ is a vector containing the coordinates of all the targets. From Fig. \ref{f1} and Fig. \ref{f2}, it can be seen that the parameter estimation performance of the beamforming scheme proposed in this paper is significantly better than that of the two linear beamforming schemes \cite{14}, zero-forcing (ZF) and maximum-ratio transmission (MRT), and is close to the ideal situation “sensing only” which is obtained without the rate constraints. When the number of antennas increases, the angle estimation performance can be better improved, since this improves the resolution of array to the angle, while the delay and positioning estimation performance in Fig. \ref{f3} are little improved, which shows the necessity of joint parameters estimation. Synthesize the above analysis, the proposed beamforming scheme can effectively reduce the dependence of the BS on the number of antennas and achieve excellent position estimation performance with low power consumption.
\begin{figure}[!h]
	\centering
%	  \vspace{-0.4cm}
	\includegraphics[width=0.45\textwidth]{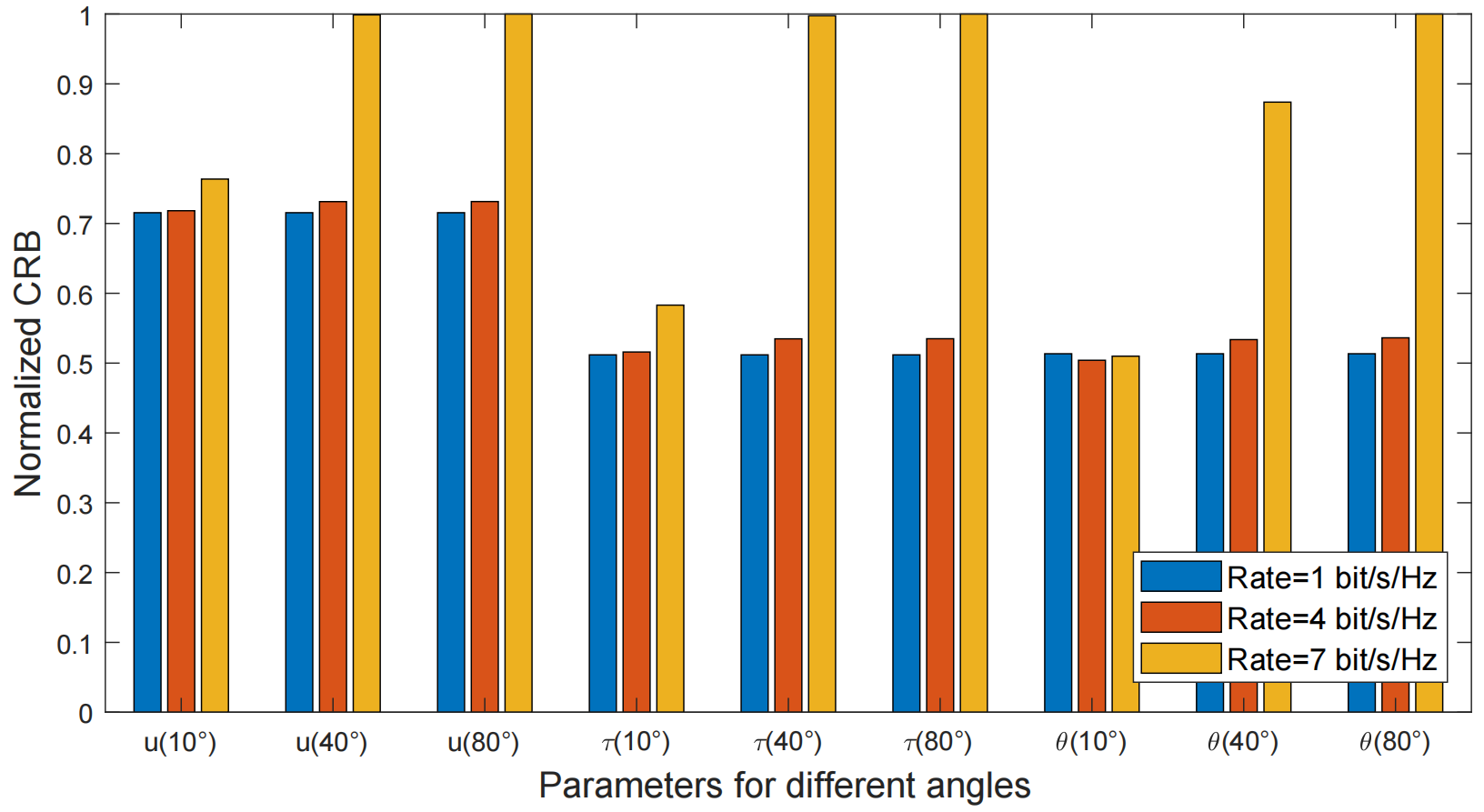}
	\captionsetup{font={footnotesize}}
	\caption{Normalized CRB of parameters at different angles, where the targets are located at azimuth angles $-30^{\circ},0^{\circ}, 15^{\circ}$, respectively, and the CUs are at azimuth angles $10^{\circ},40^{\circ}, 80^{\circ}$, respectively.}\label{f5}
	\vspace{-0.4cm}	
\end{figure}

Fig. \ref{f5} shows the impact of CUs at different locations on the compromise between sensing and communication performance.
 It can be seen that as the angle increases (meaning the distance between CU and targets increases), the same communication rate will cause more serious loss of estimation performance. It is worth noting that increasing the communication requirement in the 10° direction will improve the angle estimation performance, because the CU and the sensing target overlap in the beam energy domain, but the positioning estimation performance of the targets is still decreased, since the delay estimation performance is not improved.
\vspace{-0.1in}
\section{Conclusion}
This paper focused on the beamforming optimization problem that aims to maximize the sensing performance while adhering to communication rate and energy constraints. We
derived the CRB for joint angle and delay estimation. Specifically, we obtained the optimal beamforming for single-target and single-user case, and analysed the structure of the solution to the beamforming optimization problem over multi-target and multi-user case for computational complexity reduction. It is observed that a single parameter is insufficient to describe the positioning function adequately, prompting further investigation into the impact of CU distribution on positioning performance, which presents an intriguing avenue for future research.
\vspace{-0.1in}
\appendices
\section{Proof of Theorem \ref{CRLB}}
\label{CRLB_Derivation}
Taking the discrete Fourier transform (DFT) for the reflected data received at the BS, the frequency-domain expression of (\ref{ReflectedRadar}) can be obtained in the form \cite{9}
\begin{equation}
\begin{aligned} \label{r_f}
{\boldsymbol{r}}_{1}(\omega)\!\!=\!\!\sum_{q=1}^{Q}\sum_{k=1}^{K}\alpha_{q}{\boldsymbol{a}}(\theta_{q})b(\theta_{q})^{T}{\boldsymbol{w}}_{k}e^{-j\omega\tau_{q}}s_{k}(\omega)\!\!+\!\!{\boldsymbol{n}}_{1}(\omega), \\
\end{aligned}
\end{equation}
where $s_{k}(\omega)$ represents the DFT of $s_{k}(t)$, ${\boldsymbol{n}}_{1}(\omega) \sim \mathcal{C N}(\boldsymbol{0}, \boldsymbol{R})$ is the form of noise in frequency domain, $\boldsymbol{R} = \sigma^{2}\boldsymbol{I}_{N_r}$ \cite{9}, $\boldsymbol{I}$ represents the identity matrix. Considering the actual signal pulse shaping and calculation simplification, we set $s_{k}(\omega)=1$.
To simply notation, we denote $\boldsymbol{A}=\left[\boldsymbol{a}\left(\theta_{1}\right), \ldots, \boldsymbol{a}\left(\theta_{Q}\right)\right]$,   $\boldsymbol{B}=\left[\boldsymbol{b}\left(\theta_{1}\right), \ldots, \boldsymbol{b}\left(\theta_{Q}\right)\right]$,
 $\boldsymbol{C}=\operatorname{diag}(\boldsymbol{c})$,
 $\boldsymbol{c}=\left[\alpha_{1} e^{-j \omega \tau_{1}}, \ldots, \alpha_{Q} e^{-j \omega \tau_{Q}}\right]^{T}$,
$\boldsymbol{W}=\left[\boldsymbol{w}_{1}, \ldots, \boldsymbol{w}_{K}\right]$, 
$\boldsymbol{s}=[1, \ldots, 1]^{T}, \boldsymbol{s} \in \mathbb{C}^{K \times 1}$, then, (\ref{r_f}) can be rewritten as $\boldsymbol{r}_{1}=\boldsymbol{A C B}{ }^{T} \boldsymbol{W} \boldsymbol{s}+\boldsymbol{n}_{1}$. Using the Slepian-Bangs formula \cite{10}, for the complex observation vector $\boldsymbol{r}_1 \sim \mathcal{C N}(\boldsymbol{\mu}, \boldsymbol{R})$, in which $\boldsymbol{\mu}=\boldsymbol{A C B}{ }^{T} \boldsymbol{W} \boldsymbol{s}$, the $(m,n)$-th element of FIM $\boldsymbol{F}$ is
\begin{small} 
\begin{align}\label{FIM_ALL}
	[\boldsymbol{F}]_{m n}\!\!=\!\!\operatorname{Tr}\!\left(\!\!\boldsymbol{R}^{-1}\! \frac{\partial \boldsymbol{R}}{\partial \boldsymbol{\varsigma}_{m}} \!\boldsymbol{R}^{-1}\! \frac{\partial \boldsymbol{R}}{\partial \boldsymbol{\varsigma}_{n}}\!\right)\!\!+\!\!2\! \operatorname{Re}\!\left\{\!\operatorname{Tr}\!\left(\!\frac{\partial \boldsymbol{\mu}^{H}}{\partial \boldsymbol{\varsigma}_{m}} \!\boldsymbol{R}^{-1}\! \frac{\partial \boldsymbol{\mu}}{\partial \boldsymbol{\varsigma}_{n}}\!\right)\!\right\}.	
\end{align}
\end{small} 
Since the covariance matrix $\boldsymbol{R}$ does not depend on the parameters to be estimated, we only have to worry about the second term in (\ref{FIM_ALL}). Let $\dot{\boldsymbol{B}}\triangleq[\frac{\partial\boldsymbol{b}(\theta_1)}{\partial\theta_1},\ldots,\frac{\partial\boldsymbol{b}(\theta_{Q})}{\partial\theta_{Q}}]$ and $\dot{\boldsymbol{A}}\triangleq[\frac{\partial\boldsymbol{a}(\theta_1)}{\partial\theta_1},\ldots,\frac{\partial\boldsymbol{a}(\theta_{Q})}{\partial\theta_{Q}}]$, it follows that $\frac{\partial\boldsymbol{\mu}}{\partial{\theta}_{q}} =\dot{\boldsymbol{A}}\boldsymbol{e}_{q}\boldsymbol{e}_{q}^{T}\boldsymbol{C}\boldsymbol{B}^{T}\boldsymbol{W}\boldsymbol{s}+\boldsymbol{A}\boldsymbol{C}\boldsymbol{e}_{q}\boldsymbol{e}_{q}^{T}\dot{\boldsymbol{B}}^{T}\boldsymbol{W}\boldsymbol{s}  \label{FIM_derivation_theta}$, $\frac{\partial\boldsymbol{\mu}}{\partial\tau_{q}} =(-j\omega)\boldsymbol{A} \boldsymbol{C} \boldsymbol{e}_{q}\boldsymbol{e}_{q}^{T}\boldsymbol{B}^T\boldsymbol{W}\boldsymbol{s}, \label{FIM_derivation_tau}	
$
%\begin{align}
%	&\frac{\partial\boldsymbol{\mu}}{\partial{\theta}_{q}} =\dot{\boldsymbol{A}}\boldsymbol{e}_{q}\boldsymbol{e}_{q}^{T}\boldsymbol{C}\boldsymbol{B}^{T}\boldsymbol{W}\boldsymbol{s}+\boldsymbol{A}\boldsymbol{C}\boldsymbol{e}_{q}\boldsymbol{e}_{q}^{T}\dot{\boldsymbol{B}}^{T}\boldsymbol{W}\boldsymbol{s},  \label{FIM_derivation_theta}\\
%	&\frac{\partial\boldsymbol{\mu}}{\partial\tau_{q}} =(-j\omega)\boldsymbol{A} \boldsymbol{C} \boldsymbol{e}_{q}\boldsymbol{e}_{q}^{T}\boldsymbol{B}^T\boldsymbol{W}\boldsymbol{s}, \label{FIM_derivation_tau}	
%\end{align}
where $\boldsymbol{e}_{q}$ denotes the $q$-th column of $\boldsymbol{I}_{Q}$. Taking the submatrix $\boldsymbol{F}_{\boldsymbol{\theta}, \boldsymbol{\theta}}$ for example, then
%\begin{equation}
\begin{small} 
\begin{align}\label{FIM_theta_1}
	\left[\boldsymbol{F}_{\boldsymbol{\theta},\boldsymbol{\theta}}\right]_{p q} \!\!\!&=2\mathrm{Re}\big\{\!\!\mathrm{Tr}\big(\dot{\boldsymbol{A}}\boldsymbol{e}_{p}\boldsymbol{e}_{p}^{T}\boldsymbol{C}\boldsymbol{B}^{T}\boldsymbol{W}\boldsymbol{s}+\boldsymbol{A}\boldsymbol{C}\boldsymbol{e}_{p}\boldsymbol{e}_{p}^{T}\dot{\boldsymbol{B}}^{T}\boldsymbol{W}\boldsymbol{s}\big)^{H} \nonumber\\
	& \times\!\!\boldsymbol{R}^{-1}\big(\dot{\boldsymbol{A}}\boldsymbol{e}_q\boldsymbol{e}_q^T\boldsymbol{CB}^T\boldsymbol{W}\boldsymbol{s}\!\!+\!\!\boldsymbol{A}\boldsymbol{C}\boldsymbol{e}_q\boldsymbol{e}_q^T\dot{\boldsymbol{B}}^T\boldsymbol{W}\boldsymbol{s}\big)\!\!\big\},
\end{align}
\end{small} 
%\end{equation}
next take one of the four product terms in (\ref{FIM_theta_1}), note that
%\begin{equation}
\begin{small} 
\begin{align}\label{FIM_theta_2}
\operatorname{Tr}&\left(\left(\dot{\boldsymbol{A}} \boldsymbol{e}_{p} \boldsymbol{e}_{p}^{T} \boldsymbol{C} \boldsymbol{B}^{T} \boldsymbol{W} \boldsymbol{s}\right)^{H} \boldsymbol{R}^{-1}\left(\dot{\boldsymbol{A}} \boldsymbol{e}_{q} \boldsymbol{e}_{q}^{T} \boldsymbol{C} \boldsymbol{B}^{T} \boldsymbol{W} \boldsymbol{s}\right)\right)\notag \\
&=\boldsymbol{e}_{p}^{T}\left(\dot{\boldsymbol{A}}^{H} \boldsymbol{R}^{-1} \dot{\boldsymbol{A}}\right) \boldsymbol{e}_{q} \boldsymbol{e}_{q}^{T}\left(\boldsymbol{C} \boldsymbol{B}^{T} \boldsymbol{W} \boldsymbol{s s}^{H} \boldsymbol{W}^{H} \boldsymbol{B}^{*} \boldsymbol{C}^{H}\right) \boldsymbol{e}_{p}\notag \\
&=\left(\!\dot{\boldsymbol{A}}^{H} \boldsymbol{R}^{-1} \dot{\boldsymbol{A}}\!\right)_{pq} \left(\!\boldsymbol{C}^{*} \boldsymbol{B}^{H}\left(\boldsymbol{W} \boldsymbol{s} \boldsymbol{s}^{H} \boldsymbol{W}^{H}\right)^{*} \boldsymbol{B} \boldsymbol{C}\!\right)_{pq},
\end{align}
\end{small} 
%\end{equation}
where $\boldsymbol{X}_{pq}$ denotes $(p,q)$-th element of $\boldsymbol{X}$. The other three matrix product terms in (\ref{FIM_theta_1}) have similar forms. And $\boldsymbol{R} = \sigma^{2}\boldsymbol{I}_{N_r}$, hence, $\boldsymbol{F}_{\boldsymbol{\theta},\boldsymbol{\theta}} = \frac{2}{\sigma^{2}} \operatorname{Re}(\boldsymbol{F}^{\prime}_{\boldsymbol{\theta}, \boldsymbol{\theta}})$, with $\boldsymbol{F}^{\prime}_{\boldsymbol{\theta}, \boldsymbol{\theta}}$ given in (\ref{FIM_theta_prime}). Similar to the steps above, we can obtain the results given in (\ref{FIM_thetatau_prime}) and (\ref{FIM_tautau_prime}), respectively, in which $\odot$ represents the Hadamard (element-wise) product. As a result, the Theorem \ref{CRLB} is proved.

\section{Proof of Theorem \ref{S_D}}
\label{S_Derivation}
First of all, the optimal solution to problem (\ref{s_1}) satisfies $\boldsymbol{w}_1 \in \operatorname{span}\{\boldsymbol{b}, \dot{\boldsymbol{b}}, \boldsymbol{h}_1\}$, and the basic idea of that is to project the beamforming onto direction directly related to the target function and constraint which dominate the directions needing to be illuminated. Hence, the optimal $\boldsymbol{w}_{1}$ can be expressed as 
	\begin{equation}
		\begin{aligned}\label{s1_w1}
			\boldsymbol{w}_{1}=x_{1} \boldsymbol{d}_{x}+x_{2} \boldsymbol{d}_{y}+x_{3} \boldsymbol{d}_{z}, x_{1},x_{2},x_{3}\in \mathbb{R},
		\end{aligned}
	\end{equation}
	for $\operatorname{span}\{\boldsymbol{d}_{x}, \boldsymbol{d}_{y}, \boldsymbol{d}_{z}\}=\operatorname{span}\{\boldsymbol{b}, \dot{\boldsymbol{b}}, \boldsymbol{h}_1\}$ as well as the orthogonal basises and the objective function's constitution. What needs illustration is that $\mathrm{span}\{\boldsymbol{a}_1,\boldsymbol{a}_2,\cdots,\boldsymbol{a}_M\}$ denotes the linear space composed of the linear combination of the vectors $\boldsymbol{a}_1,\boldsymbol{a}_2,\cdots,\boldsymbol{a}_M$. It's apparent that the optimum is reached when the energy constraint gets equality causing the power budget being fully utilized, i.e., $x_{1}^{2}+x_{2}^{2}+x_{3}^{2}=P_{T}$. When the communication rate constraint
	is active, which equivalentlt means problem (\ref{s_1}) is feasible, the optimum is fulfilled when the achievable rate constraint being a equation since the power that increased the communication rate could be economized for sensing, it yields $\left|\boldsymbol{h}_{1}^{H} \boldsymbol{w}_1\right|^{2} =\left(2^{\Gamma_{1}}-1\right) \sigma_{1}^{2} $. Substitute (\ref{s1_w1}) into problem (\ref{s_1}) which can be rewritten as
	\begin{subequations}
		\begin{align} \label{s_1_w1_T}
			&\min _{\left|x_{1}\right|^{2},\left|x_{2}\right|^{2},\left|x_{3}\right|^{2}} \quad f_1 \\
			\text {s.t.}\quad& \left|x_{1}\right|^{2}\left|\boldsymbol{h}_{1}^{H} \boldsymbol{d}_{x}\right|^{2}+\left|x_{2}\right|^{2}\left|\boldsymbol{h}_{1}^{H} \boldsymbol{d}_{y}\right|^{2}+\left|x_{3}\right|^{2}\left|\boldsymbol{h}_{1}^{H} \boldsymbol{d}_{z}\right|^{2} \notag\\
			&=\left(2^{\Gamma_{1}}-1\right) \sigma_{1}^{2} \label{e1}\\
			&\left|x_{1}\right|^{2}+\left|x_{2}\right|^{2}+\left|x_{3}\right|^{2}=P_{T} \label{e2} \text {, } 
		\end{align}
	\end{subequations}
	where $f_1\!\!=\!\!\frac{\sigma^{2}}{2|\alpha_{1}|^{2}}\!\!\left(\!\frac{1}{\|\dot{\boldsymbol{a}}\|^{2}\|\boldsymbol{b}\|^{2}\left|x_{1}\right|^{2}\!+\!\|\boldsymbol{a}\|^{2}\|\dot{\boldsymbol{b}}\|^{2}\left|x_{2}\right|^{2}}\!+\!\frac{1}{\omega^{2}\|\boldsymbol{a}\|^{2}\|\boldsymbol{b}\|^{2}\left|x_{1}\right|^{2}}\!\right)\!$. For simplicity, we let $\beta_1 = \|\dot{\boldsymbol{a}}\|^{2}\|\boldsymbol{b}\|^{2}$, $\beta_2 = \|\boldsymbol{a}\|^{2}\|\dot{\boldsymbol{b}}\|^{2}$, $\beta_3 = \omega^{2}\|\boldsymbol{a}\|^{2}\|\boldsymbol{b}\|^{2}$. We first get the relationship between $\left|x_{1}\right|^{2}$ and $\left|x_{2}\right|^{2}$ through (\ref{e1})
	and (\ref{e2}), that is   $\left|x_{2}\right|^{2}=\frac{\left|\boldsymbol{h}_{1}^{H}\boldsymbol{d}_{z}\right|^{2}P_{T}-\left(2^{{\Gamma_{1}}}-1\right)\sigma_{1}^{2}}{\left|\boldsymbol{h}_{1}^{H}\boldsymbol{d}_{z}\right|^{2}-\left|\boldsymbol{h}_{1}^{H}\boldsymbol{d}_{y}\right|^{2}}-\frac{\left|\boldsymbol{h}_{1}^{H}\boldsymbol{d}_{z}\right|^{2}-\left|\boldsymbol{h}_{1}^{H}\boldsymbol{d}_{x}\right|^{2}}{\left|\boldsymbol{h}_{1}^{H}\boldsymbol{d}_{z}\right|^{2}-\left|\boldsymbol{h}_{1}^{H}\boldsymbol{d}_{y}\right|^{2}}\left|x_{1}\right|^{2} \triangleq \lambda_1-\lambda_2\left|x_{1}\right|^{2}$, according to which we can not noly calculate $x_{2}$ and $x_{3}$ after we obtain $x_{1}$ as case 1, but also obtain the range of $\left|x_{1}\right|^{2}$ that locates in $(0,\widetilde{x}_{1,\max}]$ where $\widetilde{x}_{1,\max} = \frac{\left|\boldsymbol{h}_{1}^{H} \boldsymbol{d}_{z}\right|^{2} P_{T}-\left(2^{\Gamma_{1}}-1\right) \sigma_{1}^{2}}{\left|\boldsymbol{h}_{1}^{H} \boldsymbol{d}_{z}\right|^{2}-\left|\boldsymbol{h}_{1}^{H} \boldsymbol{d}_{x}\right|^{2}}$. Substitute $\left|x_{2}\right|^{2}=\lambda_1-\lambda_2\left|x_{1}\right|^{2}$ into $f_1$, the objective function only contains one variable, which is given by $f_1(\left|x_{1}\right|^{2})\!\!=\!\!\frac{\sigma^{2}}{2|\alpha_{1}|^{2}}\!\left(\frac{1}{\beta_2\lambda_1 +(\beta_1-\beta_2\lambda_2)\left|x_{1}\right|^{2}}\!+\!\frac{1}{\beta_3\left|x_{1}\right|^{2}}\right)$, the first-order  derivate of $f_1$ with respect to $\left|x_{1}\right|^{2}$ is 
	\begin{align}
		\frac{\partial f_{1}(\left|x_{1}\right|^{2})}{\partial\left|x_{1}\right|^{2}}\!\!=\!\!\frac{\sigma^{2}}{2|\alpha_{1}|^{2}}\!\frac{g_1(\left|x_{1}\right|^{2})}{\left(\beta_2\lambda_1+(\beta_1-\beta_2\lambda_2)\left|x_{1}\right|^{2}\right)\beta_3\left|x_{1}\right|^{4}}
	\end{align}
	where $g_1(\left|x_{1}\right|^{2})=\left(\beta_2\lambda_2-\beta_1\right)\left(\beta_3+\beta_1-\beta_2\lambda_2\right)\left|x_{1}\right|^{4}+2\beta_2\lambda_1\left(\beta_2\lambda_2-\beta_1\right)\left|x_{1}\right|^{2}-\beta_2^2\lambda_1^2$.
	 Clearly, if $\beta_1-\beta_2\lambda_2 \geq 0$, the optimal solution is $\left(\left|x_{1}\right|^{2}\right)^*=\widetilde{x}_{1,\max}$ for $f_1$ decreasing monotonously with $\left|x_{1}\right|^{2}$. When $\beta_1-\beta_2\lambda_2 < 0$, there are three cases to be analysed, if $\beta_3 = \beta_2\lambda_2-\beta_1$, the only root of $g_1$ is $\frac{\beta_2\lambda_1}{2(\beta_2\lambda_2-\beta_1)}\triangleq\widetilde{x}_{1,0}$, we have $\lim _{\left|x_{1}\right|^{2} \rightarrow 0^{+}} \frac{\partial f_{1}\left(\left|x_{1}\right|^{2}\right)}{\partial\left|x_{1}\right|^{2}}<0$. Therefore, we have $\left(\left|x_{1}\right|^{2}\right)^* = \widetilde{x}_{1,0}$ if $\widetilde{x}_{1,0}\leq\widetilde{x}_{1,\max}$, and $\left(\left|x_{1}\right|^{2}\right)^*=\widetilde{x}_{1,\max}$ if $\widetilde{x}_{1,0}>\widetilde{x}_{1,\max}$. When $\beta_1-\beta_2\lambda_2 < 0$ and $\beta_3 > \beta_2\lambda_2-\beta_1$, the right root of $g_1$ is $\frac{\beta_2\lambda_1\left(\sqrt{(\beta_2\lambda_2-\beta_1)\beta_3}+\beta_1-\beta_2\lambda_2\right)}{\left(\beta_2\lambda_2-\beta_1\right)\left(\beta_3+\beta_1-\beta_2\lambda_2\right)}\triangleq\widetilde{x}_{1,\mathrm{rig1}}$, we have $\left(\left|x_{1}\right|^{2}\right)^* = \widetilde{x}_{1,\mathrm{rig1}}$ if $\widetilde{x}_{1,\mathrm{rig1}}\leq\widetilde{x}_{1,\max}$, and $\left(\left|x_{1}\right|^{2}\right)^*=\widetilde{x}_{1,\max}$ if $\widetilde{x}_{1,\mathrm{rig1}}>\widetilde{x}_{1,\max}$. When $\beta_1-\beta_2\lambda_2 < 0$ and $\beta_3 < \beta_2\lambda_2-\beta_1$, the left root of $g_1$ is $\frac{\beta_2\lambda_1\left(\sqrt{(\beta_2\lambda_2-\beta_1)\beta_3}+\beta_1-\beta_2\lambda_2\right)}{\left(\beta_2\lambda_2-\beta_1\right)\left(\beta_3+\beta_1-\beta_2\lambda_2\right)}\triangleq\widetilde{x}_{1,\mathrm{lef2}}$ and the right root of $g_1$ is $\frac{\beta_2\lambda_1\left(-\sqrt{(\beta_2\lambda_2-\beta_1)\beta_3}+\beta_1-\beta_2\lambda_2\right)}{\left(\beta_2\lambda_2-\beta_1\right)\left(\beta_3+\beta_1-\beta_2\lambda_2\right)}\triangleq\widetilde{x}_{1,\mathrm{rig2}}$. Since $g_1 < 0$ when $\left|x_{1}\right|^2 \in (0,\widetilde{x}_{1,\mathrm{lef2}}) \cup (\widetilde{x}_{1,\mathrm{rig2}},+\infty)$ and $g_1 > 0$ when $\left|x_{1}\right|^2 \in (\widetilde{x}_{1,\mathrm{lef2}},\widetilde{x}_{1,\mathrm{rig2}})$, we have $\left(\left|x_{1}\right|^{2}\right)^* = \widetilde{x}_{1,\max}$ if $\widetilde{x}_{1,\mathrm{lef2}}\geq\widetilde{x}_{1,\max}$, and $\left(\left|x_{1}\right|^{2}\right)^*=\widetilde{x}_{1,\mathrm{lef2}}$ if $\widetilde{x}_{1,\mathrm{lef2}}<\widetilde{x}_{1,\max}\leq\widetilde{x}_{1,\mathrm{rig2}}$. If $\widetilde{x}_{1,\mathrm{rig2}} < \widetilde{x}_{1,\max}$, we need compare $f_1(\widetilde{x}_{1,\mathrm{lef2}})$ with $f_1(\widetilde{x}_{1,\max})$, the optimal solution between $\widetilde{x}_{1,\mathrm{lef2}}$ and $\widetilde{x}_{1,\max}$ is determined by the smaller objective value. This results in the solutions in Theorem \ref{S_D}, which completes the proof.
\section{Proof of Proposition \ref{M_D}}
\label{M_Derivation}
We know that FIM and $\boldsymbol{W}_{v}$ are connected by $\boldsymbol{B}$ and $\dot{\boldsymbol{B}}$ on account of Theorem \ref{CRLB}. Decompose $\boldsymbol{W}_{v}^{*}=\boldsymbol{\Omega}_{U_{T}} \boldsymbol{W}_{v}^{*}+\boldsymbol{\Omega}_{U_{T}}^{\perp} \boldsymbol{W}_{v}^{*}$, in which $\boldsymbol{\Omega}_{U_{T}}$ denotes the orthogonal projection onto the subspace spanned by the columns of $\boldsymbol{U}_{T}$ in (\ref{M_W}) and $\boldsymbol{\Omega}_{U_{T}}^{\perp}=\boldsymbol{I}-\boldsymbol{\Omega}_{U_{T}}$.
According to the analysis of \cite{CRBLJ}, the columns of the optimal $\boldsymbol{W}_{v}^{*}$  belong to the subspace spanned by the columns of $\boldsymbol{U}_{T}$, that is to say, $\boldsymbol{\Omega}_{U_{T}}^{\perp} \boldsymbol{W}_{v}^{*}=0$. Hense, the optimal beamforming covariance matrix can be given by 
\begin{small} 
\begin{align}
		\boldsymbol{W}_{v}^{*}=\boldsymbol{\Omega}_{\boldsymbol{U}_{T}} \boldsymbol{w}_{v} \boldsymbol{w}_{v}^{H} \boldsymbol{\Omega}_{\boldsymbol{U}_{T}}=\boldsymbol{U}_{T} \boldsymbol{\Lambda} \boldsymbol{U}_{T}^{H},
%	\boldsymbol{\Omega}_{U_{T}}=\boldsymbol{U}_{T}\left(\boldsymbol{U}_{T}^{H} \boldsymbol{U}_{T}\right)^{-1} \boldsymbol{U}_{T}^{H}, \\
\end{align}
\end{small} 
where $\boldsymbol{\Lambda}=\left(\boldsymbol{U}_{T}^{H} \boldsymbol{U}_{T}\right)^{-1} \boldsymbol{U}_{T}^{H} \boldsymbol{w}_{v} \boldsymbol{w}_{v}^{H} \boldsymbol{U}_{T}\left(\boldsymbol{U}_{T}^{H} \boldsymbol{U}_{T}\right)^{-1}$.

\bibliographystyle{ieeetr}
%\IEEEtriggeratref{1} % 表示对第35个参考文献开始换栏
\balance
\bibliography{ref}
\end{document}